\documentstyle[aps,psfig]{revtex}
\begin{document}

\title{Corrections to scaling at the Anderson transition}

\author{Keith Slevin}
\address{Dept. of Physics, Graduate School of Science,
Osaka University, \\ 1-16 Machikaneyama, Toyonaka,
Osaka 560, Japan}

\author{Tomi Ohtsuki}

\address{Department of Physics, Sophia University,
Kioi-cho 7-1, Chiyoda-ku, Tokyo 102-8554, Japan}
\maketitle

\begin{abstract}
We report a numerical analysis of corrections to finite size scaling
at the Anderson transition due to irrelevant scaling variables and
non-linearities of the scaling variables.
By taking proper account of these corrections, the
universality of the critical exponent for the orthogonal
universality class for three
different distributions of the random potential is convincingly
demonstrated.
\end{abstract}


\twocolumn

The possibility of the Anderson localization of electron
states as a result of disorder was first suggested
four decades ago \cite{ANDERSON}.
Following the proposal of the scaling theory of
localization \cite{SCALING} attention has focused on understanding
the critical properties of the Anderson transition (AT),
the quantum phase transition
which occurs at a critical disorder separating a diffusive
metallic phase from an insulating localized phase \cite{KM}.

Our current understanding of the AT is based on the
non-linear $\sigma$ model (NL$\sigma$M) \cite{EFETOV}.
This has been analyzed using an expansion
in powers of $\epsilon$, where $d=2+\epsilon$ is the dimension
of the system.
According to the NL$\sigma$M it should be possible to classify
the critical behavior using three universality classes:
orthogonal, unitary and symplectic depending on the symmetry
of the Hamiltonian with respect to time reversal and spin rotation.
Here we focus on the orthogonal universality class
corresponding to systems with both time reversal and spin
rotation symmetries.

Beyond the suggestion of the appropriate universality classes,
there has not been much success in making
detailed predictions about the critical behavior with the NL$\sigma$M.
The problems are well illustrated by attempts to estimate
the critical exponent $\nu$ which describes the divergence
of the correlation length $\xi$ at the AT.
In early work it was found that $\nu=1/\epsilon$ \cite{hikami2} which
gives $\nu=1$ when extrapolated to $d\equiv \epsilon+2 =3$.
When combined with the Wegner scaling law $s=\nu \epsilon$ \cite{WEGNER}
this leads to a conductivity exponent $s=1$.
Measurements on some, but not all, materials do indeed yield $s=1$
\cite{ITOH}.
However, calculations at higher orders in $\epsilon$ produced strong
corrections to the leading order when extrapolated to $\epsilon=1$
\cite{WEGNER2,HIKAMI}, showing that that this agreement is fortuitous.
There is now no accepted estimate of the exponent based on the $\epsilon$
expansion or any other analytic technique.

While the above can be regarded as a rather unfortunate technical
difficulty, fears have also been expressed that there maybe an
infinite number of relevant operators in the NL$\sigma$M \cite{KRAV}
and that the theory may be unsound.
While it now seems unlikely in view of \cite{BREZIN}
that this is actually the case, in this context it is
nevertheless important
to have independent confirmation that our understanding of the AT
is correct.
At present numerical simulations \cite{KM,MAC,SHK,ZHAR} offer the only viable
alternative.

In this paper we demonstrate an important basic principle underlying our
understanding of the AT: the universality of the critical properties
of the AT.
To do this has required us to address the principle uncertainty in previous
numerical studies of the critical properties of the AT,
the presence of systematic corrections to scaling in the numerical data
due to the practical limitations on the sizes of system which can be studied.

The computer time required in numerical studies of the AT increases
very rapidly with increasing system size (as $L^7$ for the method used here.)
This sets a severe limitation on the system sizes which can be simulated.
However, systematic corrections to scaling are expected in smaller systems
and their
neglect leaves important question marks over the validity
of any conclusions drawn from the analysis of the numerical data.
Here we consider two ways in which such corrections
can arise: the presence of irrelevant scaling variables
and non-linearity of the scaling variables \cite{CARDY}.
These effects lead to systematic rather than random
deviations from scaling and
must be taken into account both when estimating the
critical parameters and the likely
accuracy of their estimation.

Our work has also been inspired by the
successful analyses of corrections to scaling in
the Quantum Hall Effect (QHE) transition \cite{HUCKE,WANG}.
The present problem is, however,
more difficult since, unlike the QHE, the critical point
is not known {\it a priori} on grounds of symmetry.

The universality of the exponent for the box and Gaussian distributions
of random potential was demonstrated to a limited extent in \cite{MAC}
by taking account of corrections to scaling in an ad hoc manner.
Here, taking account of corrections systematically, we confirm
that result and extend its validity to include the Lloyd
model \cite{THOULESS}.

The Hamiltonian used in this study describes
non- interacting electrons on a simple cubic lattice
with nearest neighbor interactions only
\[
\begin{array}{lll}
<\vec r| H | \vec r> & =  & V(\vec r) , \\
<\vec r| H | \vec r - \hat x> & =    & -1 , \\
<\vec r| H | \vec r - \hat y> & =    & -1 , \\
<\vec r| H | \vec r - \hat z> & =    & -1 .
\end{array}
\]
Here $\hat x$, $\hat y$ and $\hat z$ are the lattice basis vectors.
The potential $V$ is independently and identically distributed
with probability $p(V)\text{d}V$.
We studied three models of the potential distribution:
The box distribution
\[
\begin{array}{llll}
p(V) & = &1/W & |V| \leq W/2  ,\\
                            & = & 0 & \mathrm{otherwise} ,
\end{array}
\]
the Gaussian distribution
\[
p(V) = \frac{1}{\sqrt{2 \pi \sigma^2}} \exp\left( - \frac{V^2}{2 \sigma^2}
\right) ,
\]
with $\sigma^2=W^2/12$,
and the Lloyd model in which $V$ has a Lorentz distribution
\[
p(V) = \frac{W}{\pi \left( W^2 + V^2 \right)} .
\]
For this distribution all moments higher than the mean
are divergent and the
parameter $W$ is proportional to the
full width at half maximum
of the distribution.
For these three models we analyzed the finite size scaling
\cite{CARDY}
of the localization length $\lambda$ for electrons on a
quasi-$1d$ dimensional bar of cross section $L \times L$.
The length $\lambda$ was determined to within
a specified accuracy using a
standard transfer matrix technique \cite{PS,MK}.

The starting point of our analysis is the renormalization
group equation which expresses the dimensionless quantity
$\Lambda=\lambda/L$ as function of the scaling variables
\[
\Lambda=f\left( \frac{L}{b}, \chi b^{1/\nu}, \psi b^{y}
\right) .
\]
In this equation $b$ is the scale factor
in the renormalization group, $\chi$ the relevant scaling variable
and $\psi$ the leading irrelevant scaling variable.
We should find $ y<0$ if
$\psi$ is irrelevant.
An appropriate choice of the factor $b$ \cite{CARDY}
leads to
\begin{equation}
\Lambda=F\left( \chi L^{1/\nu}, \psi
L^{y}\right)
\label{fit0} ,
\end{equation}
where $F$ is a function related to $f$.

For $L$ finite there is no phase transition and
$F$ is a smooth function of its arguments.
Assuming the irrelevant scaling variable is not
dangerous, we make a Taylor expansion up to order $n_I$
\begin{equation}
\Lambda=\sum_{n=0}^{n_I} \psi^n L^{n y}
F_n\left( \chi L^{1/\nu}\right) ,
\label{fit1}
\end{equation}
and obtain a series of functions $F_n$. Each $F_n$ is
then expanded as a Taylor series up to
order $n_R$
\begin{equation}
F_n(\chi L^{1/\nu}) =  \sum_{m=0}^{n_R} \chi^m L^{m/\nu} F_{nm}.
\label{fit2}
\end{equation}
To take account of non- linearities in the
scaling variables we expand both in terms of the dimensionless
disorder $w=(W_c-W)/W_c$
where $W_c$ is the critical disorder separating the
insulating $(w<0)$ and conducting phases $(w>0)$.
\begin{equation}
\begin{array}{ccc}
\chi(w) = \sum_{n=1}^{m_R} b_n  w^n  & , &
\psi(w) = \sum_{n=0}^{m_I} c_n  w^n .
\end{array}
\label{fit4}
\end{equation}
The orders of the expansions are
$m_R$ and $m_I$ respectively.
Notice that $\chi(0)=0$.
The absolute scales of the arguments in (\ref{fit0}) are
undefined, we fix them by setting $F_{01}=F_{10}=1$ in
(\ref{fit2}).
The total number of fitting parameters is $N_p= (n_I+1)(n_R+1)+ m_R +m_I +2$

The qualitative nature of the corrections
can be understood by looking at some special cases.
First let us suppose that non-linearities are absent
($m_R=1$ and $m_I=0$) and truncate (\ref{fit1}) at $n_I=1$
\[
\Lambda=F_0\left(\chi L^{1/\nu}\right) +
\psi L^{y}F_1\left(\chi L^{1/\nu}\right) .
\]
From this equation we can infer that the estimate of
the critical disorder, and possibly also the critical exponent,
will appear to shift in a systematic way as the
size of the system increases.
To exhibit scaling it is necessary to subtract the corrections
due to the irrelevant scaling variable. When $n_I=1$ we define
\begin{equation}
\Lambda_{\mathrm{corrected}} = \Lambda -
\psi L^{y}F_1\left(\chi L^{1/\nu}\right) ,
\label{corrected}
\end{equation}
with the obvious generalization when $n_I>1$. We then have
\begin{equation}
\Lambda_{\mathrm{corrected}}=F_{\pm}\left(\frac{L}{\xi}\right).
\label{scalfunc}
\end{equation}
The functions $F_{\pm}$ are defined by
$F_{\pm}(x)=F_0(\pm (\xi_\pm x)^{1/\nu}).$
In this case the correlation length $\xi$ has a simple
power law dependence on the dimensionless disorder
$\xi=\xi_{\pm} \left| w \right|^{-\nu}$.
The constants $\xi_{\pm}$ are not normally determined in
finite size scaling studies.

On the other hand, if we neglect the irrelevant variable and
consider only non-linearity in the scaling variable we find
$\Lambda=F_{\pm}\left(L/\xi\right)$ without the need to subtract any
corrections.
No systematic shift of the estimated critical point
should occur as the system size is increased.
However, the correlation length $\xi$ no longer has a simple
power law dependence on $w$ but
behaves as $\xi=\xi_{\pm} \left| \chi \right|^{-\nu}$.

The critical exponent $\nu$, the irrelevant exponent $y$
and the functions $F_n$ are expected to be universal, while
the coefficients $\{ b_n\}$ and $\{ c_n\}$ are not.
Though we have explicitly considered
corrections due to the leading irrelevant scaling variable only,
the analysis can be easily extended to several such variables.

In the simulation $\lambda$ was evaluated as function
of disorder $W$ for a range of system sizes $L$.
The best fit was determined by
minimizing the $\chi^2$ statistic \cite{SIVIA}.
This is justified if we suppose a uniform prior
probability for all parameters, that the deviations
between the model and the simulation data are
purely random in origin and distributed
following a Gaussian distribution \cite{SIVIA}.
This last assumption is also important in
determining the likely accuracy to which the critical
parameters have been estimated.
Since the inclusion of corrections to scaling allows for
systematic rather than just random deviations from scaling
in the numerical data, this
assumption is more reasonable here
than when corrections to scaling are neglected.
Therefore we expect the estimates of the accuracy of the
critical exponent etc. to be more reliable.
The model (\ref{fit1})-(\ref{fit4}) is nonlinear in some
parameters so the goodness of fit $Q$ has been checked
using a Monte Carlo technique and the confidence intervals evaluated by
re-sampling \cite{NUMREP,DAVISON}.

The inclusion of the corrections
in (\ref{fit1})-(\ref{fit4}) leads
to a rapid increase in the number of fitting parameters and
high quality data are essential if meaningful
fits are to be obtained. All
data used here have an accuracy of either $0.1\%$
or $0.05\%$.
To achieve this accuracy between $10^6$ and $10^7$
iterations in the transfer matrix method were required.
When deciding which correction terms to include
we attempted to maximize
the goodness of fit $Q$ while keeping the number of correction
terms to a minimum.

The details of the simulations and the types of fit used
are listed in Table \ref{TPARAMS}.
The estimated critical parameters and
their confidence intervals are given in Table \ref{TBESTFIT}.
Some typical data are displayed in Fig. \ref{F1}.
To exhibit scaling the data are re-plotted after
subtraction of the appropriate corrections in Fig. \ref{F2}.
The corrected data now fall on a single curve clearly
exhibiting scaling in agreement with (\ref{scalfunc}).
The magnitude of the corrections needed to obtain the scaling
shown in Fig. \ref{F2} are of the order of $2\%$ or so for
the smallest system size decreasing to around $0.3\%$
for the largest system size.

The most important point to be drawn from Table \ref{TBESTFIT} is
that the estimates of the exponent $\nu$ for the three different
disorder distributions are in almost perfect agreement.
The same is true for the estimates of the
critical parameter $\Lambda_c$.
This is strong evidence in favor of the universality
of the critical exponent and other critical parameters.

How do the results of the present analysis compare
with those obtained when corrections to scaling are neglected?
In Table \ref{TNOCORR} we give the results obtained for
the same potentials neglecting corrections.
The first thing to notice is that range of system sizes (and in the box
distribution, the range of $W$) for
which an acceptable fit ($Q>0.1$) can be achieved is very limited.
After discarding data for the smaller system sizes reasonable
agreement is obtained between Tables \ref{TBESTFIT} and \ref{TNOCORR}.
However, given the more limited range of system sizes, the
estimates of the accuracy to which the critical parameters have
been determined when corrections are neglected
are too optimistic.
The problem is more evident when looking at
less accurate e.g. $0.2\%$ data \cite{SLEVIN} for the box distribution.
Ignoring corrections to scaling, it
was found that $W_c=16.45\pm.01$, $\Lambda_c=0.586\pm.001$ and
$\nu = 1.59\pm 0.03$.
The estimates of $W_c$ and $\Lambda_c$ are not consistent with 
Table \ref{TBESTFIT}.
The effect which gives rise to this inconsistency can also
be seen in the data for Lloyd model displayed in Fig. \ref{F1}.
A systematic shift of the apparent critical disorder to a lower value
as the system size increases is evident.
For the box and Gaussian distributions the shift was found to
be in the opposite sense to higher disorder.
It seems likely that any analysis which assumes that deviations from
scaling are purely random origin, rather than allowing for systematic
corrections such as considered here, will lead to an
over optimistic estimate of the accuracy to which the critical point has been
determined and even to an incorrect determination of the
critical point.
In contrast the estimate of the critical
exponent is quite consistent with that in Table \ref{TBESTFIT}.
Of course, the precise location of the critical point in any
particular model is not in itself very important,
but any inaccuracy in its estimate
also affects the estimate of the
critical conductance distribution \cite{SLEVIN}.

We should also mention that
surface effects and
the influence of boundary conditions
may also give rise to corrections to scaling behavior.
We have used periodic boundary conditions to
minimize surface effects.
Even so there may remain some influence of the boundary
conditions.
To quantify this we have evaluated $\lambda$ for
some representative values of the parameters
for fixed (fbc), periodic (pbc)  and anti- periodic boundary conditions
(apbc).
A large statistically significant shift in the localization
length $\lambda$ was found between fbc and pbc.
However, even when calculating at a higher accuracy of $0.02\%$
no such difference between pbc and apbc was found.
We therefore think it reasonable to neglect such corrections
here.

We have presented a numerical study of the Anderson transition
in three dimensions in which systematic corrections to scaling have been
explicitly taken into account when estimating the critical disorder
and other critical parameters.
The universality of the critical exponent with respect to
the choice of the distribution of disorder has been accurately
verified.

While in this paper we concentrated on the scaling of the
correlation length in the three dimensional Anderson model,
corrections of a similar nature may also be important in
finite size scaling analyses of
the conductance distribution and energy level statistics.
The method we have described here is applicable in these cases and,
indeed, to any continuous quantum phase transition.

Part of this work has been carried out on supercomputer facilities at
the Institute for Solid State Physics, University of Tokyo.

\begin{table}
\begin{tabular}{|l|l|l|l|l|l|l|l|l|l|}
Disorder & $n_R$ & $n_I$ & $m_R$ & $m_I$ &
 $W$ & $N_d$ & $N_p$ & $\chi^2$ & $Q$ \\ \hline
Box      & 3 & 1 & 2 & 0 & [15,18] & 224 & 12 & 214 & 0.5 \\
Gaussian & 2 & 1 & 2 & 0 & [20.2,22.2] & 175 & 10 & 174 & 0.3 \\
Lorentz  & 2 & 2 & 1 & 0 & [4.1,4.5] & 224 & 12 & 203 & 0.7 \\
\end{tabular}
\caption{The disorder distribution, the type of fit,
the range of disorder $W$,
the number of data $N_d$, the number of parameters $N_p$,
the value of $\chi^2$ for the best fit and
goodness of fit $Q$. The system sizes used were
$L=4,5,6,8,10,12,14$.}
\label{TPARAMS}
\end{table}

\begin{table}
\begin{tabular}{|l|l|l|l|l|}
  & $W_c$          & $\Lambda_c$    &  $\nu$        &
$y$\\ \hline
B  & 16.54(53,56) & 0.576(74,78) & 1.57(55,59) & -2.8(3.3,2.3)\\
G  & 21.29(28,31) & 0.576(74,77) & 1.58(55,61) & -3.9(5.9,2.7)\\
L  & 4.265(52,72) & 0.579(76,88) & 1.58(47,65) & -2.5(3.2,1.3)\\
\end{tabular}
\caption{The best fit estimates of the critical disorder
and the critical
exponent and their $95\%$ confidence intervals.
The quantity $\Lambda_c= F_0(0)$ is expected to be universal.}
\label{TBESTFIT}
\end{table}

\begin{table}
\begin{tabular}{|l|l|l|l|l|l|}
  & $L$      &     $W$  & $W_c$           & $\Lambda_c$    &  $\nu$   \\ \hline
B & $\ge8$   & [16,17]  & $16.514(07,22)$ & $0.579(78,80)$ & $1.58(53,63)$ \\
G & $\ge8$   & [20.2,22.2]  & $21.28(26,29)$ & $0.577(76,78)$ &
$1.58(54,62)$ \\
L & $\ge10$   & [4.1,4.5]  & $4.275(72,78)$ & $0.574(73,75)$ & $1.58(53,62)$\\
\end{tabular}
\caption{Best estimates of the critical parameters when corrections
 to scaling are neglected.}
\label{TNOCORR}
\end{table}

\begin{figure}
\caption{$\Lambda$ as a function of disorder
for the three dimensional Lloyd model.
The solid lines are the fit (\ref{fit1})-(\ref{fit4}).}
\label{F1}
\end{figure}

\begin{figure}[b]
\caption{The data in Fig. 1 after subtraction of corrections to scaling
(see (\ref{corrected})) together with the the scaling functions (\ref{scalfunc}).
Here $\xi=\xi_{\pm}\left|\chi \right|^{-\nu}$.
The upper branch corresponds to the metallic phase and 
the lower branch to the insulating phase.}
\label{F2}
\end{figure}

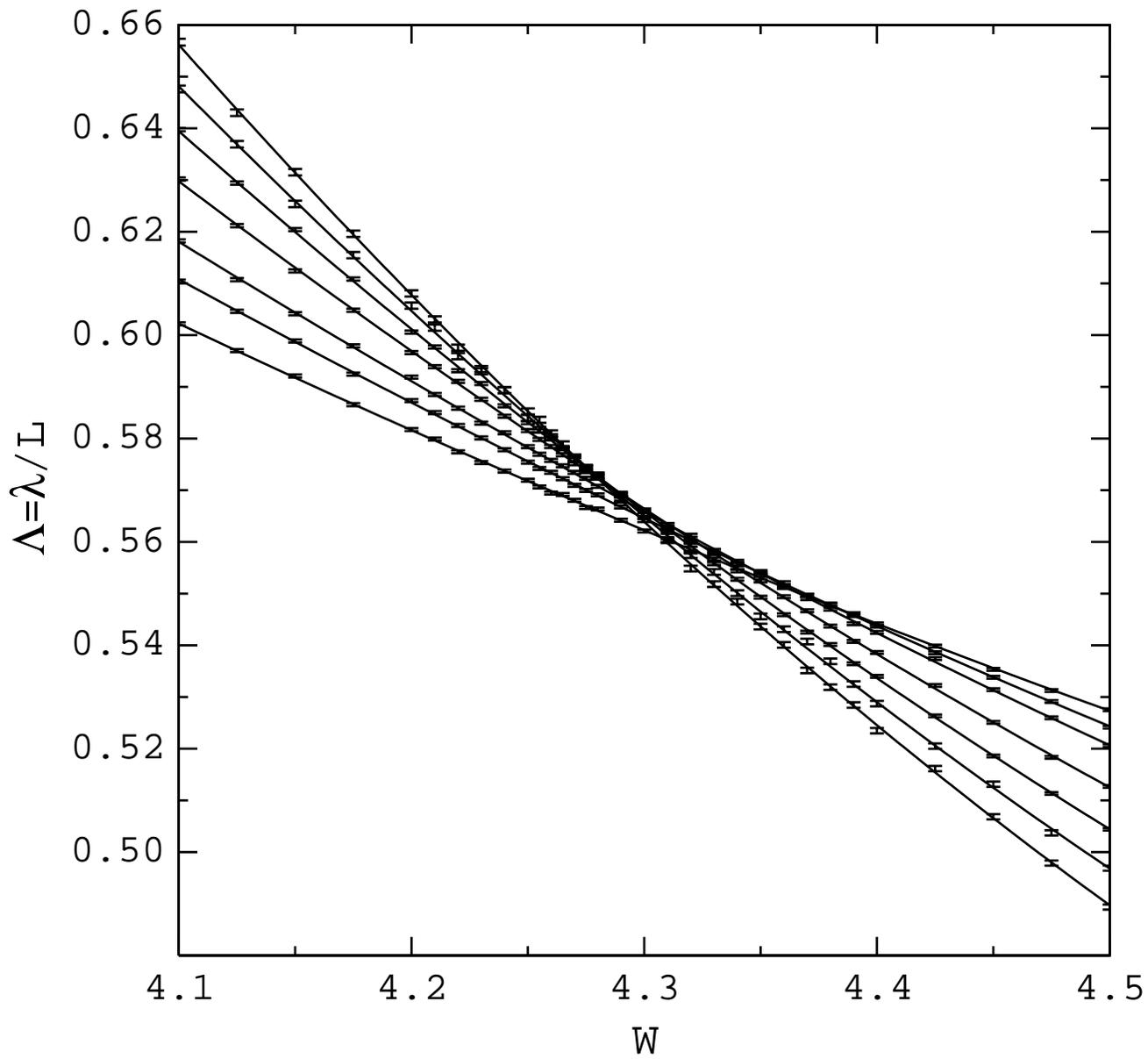
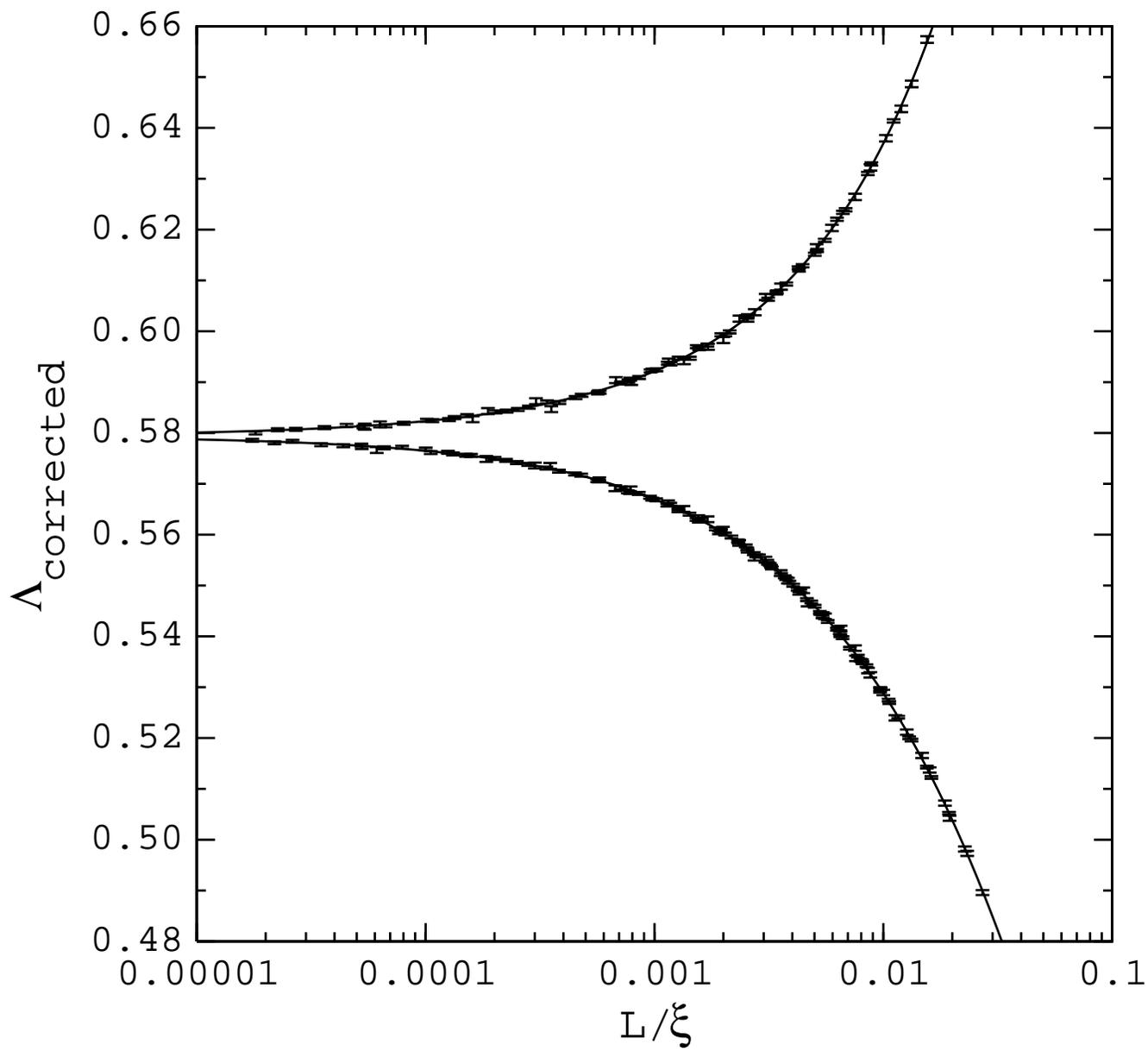
\end{document}